\documentclass[11pt]{article}
\usepackage{moriond,epsfig}
\usepackage{amssymb,amsmath}

\def\be{\begin{equation}}
\def\ee{\end{equation}}
\def\bea{\begin{eqnarray}}
\def\eea{\end{eqnarray}}

\begin{document}
\vspace*{4cm}
\title{Jet quenching}

\author{ Carlos A. Salgado}

\address{Department of Physics, CERN, Theory Division, CH-1211 Gen\`eve 23,
Switzerland}

\maketitle\abstracts{
High-$p_t$ particles produced in nucleus-nucleus collisions constitute a
powerful tool to study the medium properties. The energy loss resulting from
the propagation of these particles in the produced medium 
translates into a suppression of the
high-$p_t$ yields. These effects are usually associated to medium-induced
gluon radiation which, in turn, predicts a broadening of the jet-like signals.
Both the energy loss and the jet broadening are expected to increase
proportionally to the medium density. In the more realistic case of a
dynamically expanding medium, the gluon radiation becomes anisotropic due to
the presence of a preferred direction in the transverse plane with respect to
the axis of propagation. This anisotropy translates into deformed jet-shapes
which provide new posibilities to study these flows by high-$p_t$
measurements.
}

\section{Introduction}

Recent results 
at high transverse momentum from RHIC\cite{exp} 
on the inclusive particle suppression 
and the absence of away-side two particle correlations in central heavy
ion collisions, together with the negative-effect results from the reference
deuteron-gold run, suggest the formation of a very dense partonic medium with
which the triggered particles interact strongly. The nature of this medium is
still unknown and the study of its properties is the main goal of the
experimental program of high-energy heavy ion collisions. More differential
measurements of particles with high transverse momentum will give very valuable 
information as explained in the next sections.

In heavy ion collisions, the particles produced perturbatively at high 
transverse momentum are expected to be uncorrelated from the small transverse
momentum {\it bulk}. At the same time, this high-multiplicity state is 
expected to form a deconfined and thermalized medium which modifies the 
properties of the parton shower developed by the high-$p_t$ particles. The
formalism to compute the medium-induced gluon radiation has been developed
using several techniques and different approximations \cite{reviews}. 
Apart from details,
most of the main results depend on coherence effects which suppress gluon
radiation at small transverse momentum and/or large energies $k_t^2 \lesssim 
\hat q L$, $\omega \gtrsim \hat q L^2$, where the transport coefficient 
$\hat q$ depends on properties of the medium as the density. This results in
the well-known quadratic dependence of the radiative energy loss with the
traverse length of the medium. Moreover, the medium-modification of the
transverse momentum spectrum of radiated gluons translates into a jet 
broadening $\langle k_t\rangle\sim \Delta E\, L/\alpha_s$ \cite{Baier:1996sk}. 
One of the main
predictions of this formalism is, then, the broadening of the associated gluon
radiation when compared with the evolution in the vacuum.

\section{Energy loss and jet quenching}

The medium-induced gluon radiation spectrum $\omega dI/d\omega$ depends on the
length of the medium and the transport coefficient $\hat q$. In the absence of
a more elaborated formalism, taking into account interference effects on
multiple gluon radiation, the independent gluon emission approximation is
usually taken. In this way, the probability that an additional energy $\Delta
E$ is radiated by medium effect is given by the {\it quenching weights}
\cite{Baier:2001yt,Salgado:2003gb} 
$P(\Delta E)$ and the medium-modified fragmentation functions are modeled by
the convolution $D^{\rm med}_{i\to h}=P(\Delta E)\otimes
D_{i\to h}$. These medium-modified fragmentation functions can be used to
compute the cross section for the production of a hadron $h$ through the
perturbative expression [for precise definitions of the convolutions
see e.g. \cite{Eskola:2004cr}]
\begin{equation}
\frac{d\sigma^{AA\to h}}{dp_t}\sim f_i^A(x_1,Q^2)\otimes f_j^A(x_2,Q^2)
\otimes \sigma^{ij\to k}\otimes D^{\rm med}_{i\to h}(z,\mu_F)
\end{equation}
The strategy is then to fit the best value of $\hat q$ that reproduces the
experimental suppression measured by the ratio
\begin{equation}
R_{AA}(p_t)=\frac{dN^{AA}/dp_t}{N_{\rm coll}dN^{pp}/dp_t}.
\end{equation}
The transport coefficient $\hat q$ is proportional to the medium density. 
Thus, comparing the different $\hat q$ obtained by applying this procedure to 
different
systems, information about the density of the media is obtained. One
limitation, however, appears when the medium is very dense and the suppression
is so strong that the effect is dominated by surface emission 
\cite{Eskola:2004cr,Muller:2002fa}. 
In this case,
the measure gives only a lower limit for the transport coefficient
\cite{Eskola:2004cr}. In order 
to improve the determination of the medium properties, one possible solution
is to study the case of heavy quarks \cite{Armesto:2005iq}. 
In this case, the radiation is suppressed by mass terms \cite{hqeloss}
and, hence, the effect is smaller. In Fig.1
the prediction \cite{Armesto:2005iq}
for the suppression of electrons from the decay of charm
quarks at RHIC is presented using the transport coefficient obtained from the
 light meson case.

\begin{figure}
\begin{minipage}{0.5\textwidth}
\includegraphics[width=0.9\textwidth]{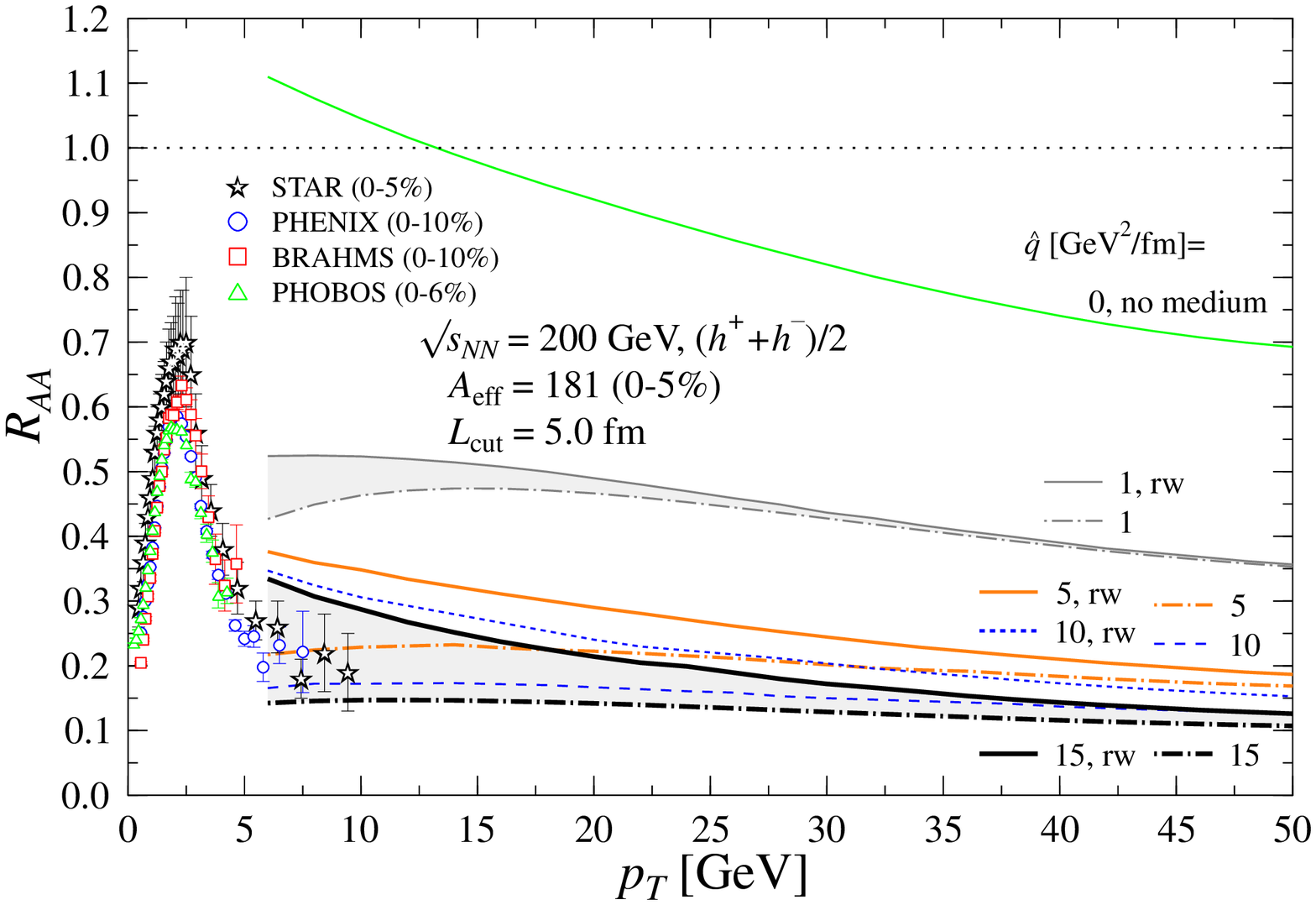}
\end{minipage}
\hfill
\begin{minipage}{0.5\textwidth}
\includegraphics[width=0.9\textwidth]{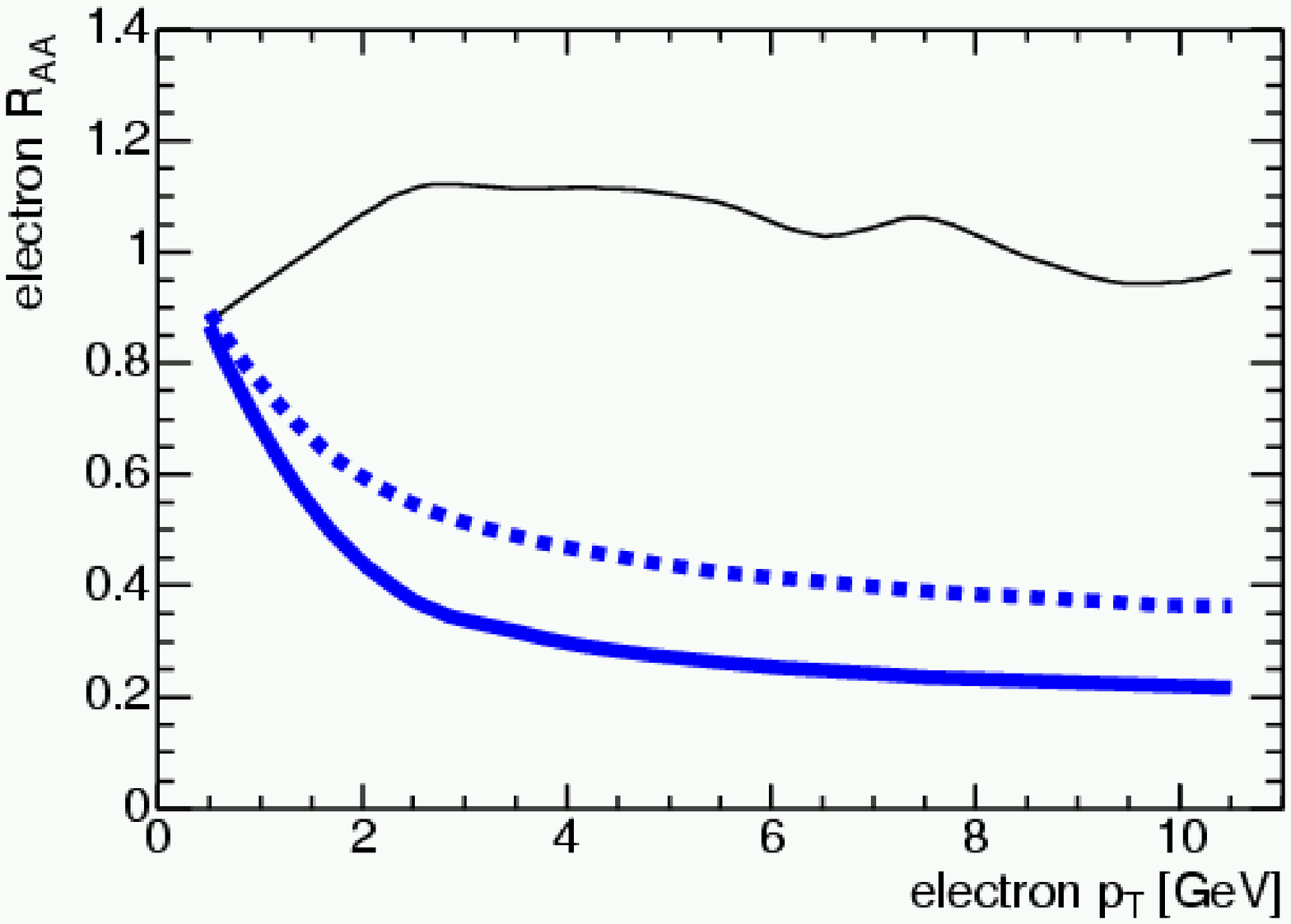}
\end{minipage}
\caption{Left:
nuclear modification factor $R_{AA}$ for charged particles in central AuAu
collisions at $\sqrt{s}=200$ GeV compared with theoretical curves
~\protect\cite{Eskola:2004cr} for different values of $\hat q$. Right:
Prediction ~\protect\cite{Armesto:2005iq}
for the suppression of electrons coming from the decay of charm
quarks in central AuAu collisions at $\sqrt{s}=200$ GeV.}
\end{figure}

\section{Medium-modified jet shapes}

The structure of the jets is expected to be strongly modified when developed
in a medium. The larger emission angle of the medium-induced spectrum translates
into a broadening of the jet shapes. Although the broadening in energy could
remain small for moderate transport coefficients, 
the intrajet multiplicity 
distribution is expected to present a harder spectrum in the transverse momentum
with respect to the jet axis \cite{Salgado:2003rv}
(see Fig. 2). This situation would be ideal for 
the study of medium-modified jet shapes at the LHC as i) it would allow for a 
good 
calibration of the jet energy (essential in order to study the jet properties) 
for the moderate values of the jet cone ($R\sim 0.3$) to be measured in the
high-multiplicity environment of a heavy ion collision; and ii) the broadening 
produced in the intrajet multiplicities would be sizable enough to measure
with high precision the medium effects.

\begin{figure}
\begin{minipage}{0.5\textwidth}
\begin{center}
\includegraphics[width=0.8\textwidth]{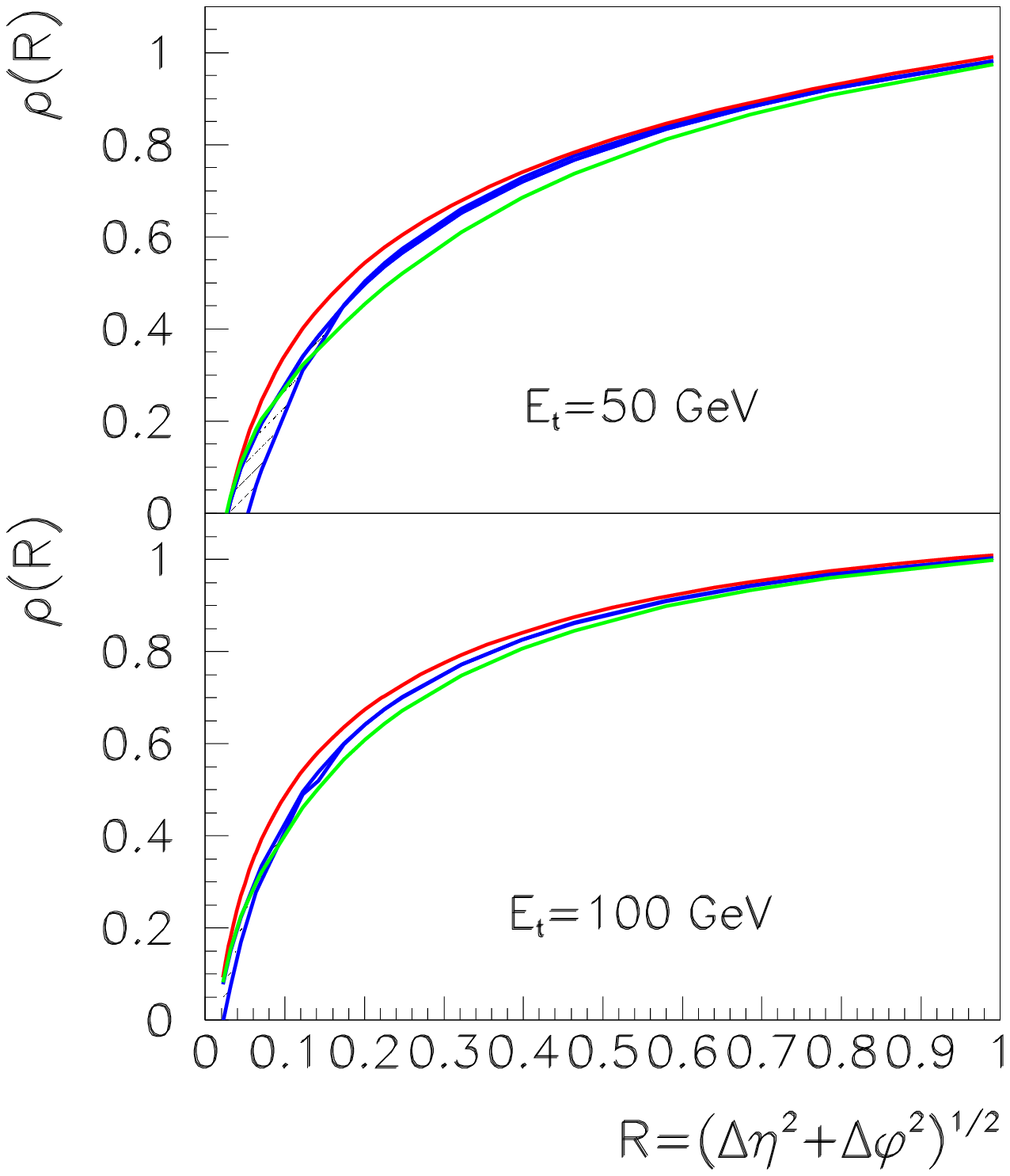}
\end{center}
\end{minipage}
\hfill
\begin{minipage}{0.5\textwidth}
\begin{center}
\includegraphics[width=0.8\textwidth]{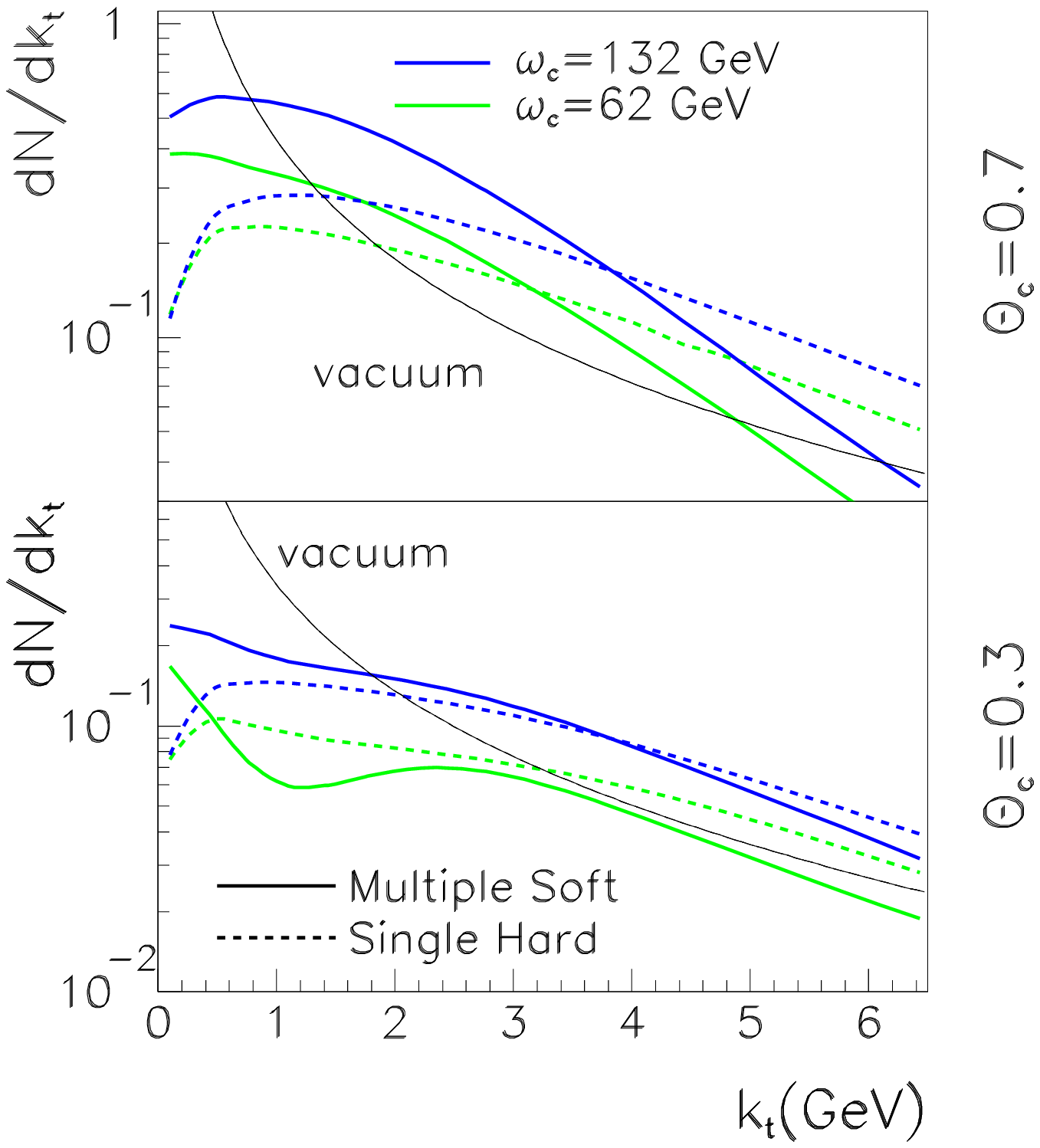}
\end{center}
\end{minipage}
\caption{Left: Fraction of the jet energy inside a cone 
$R=\sqrt{\Delta \eta^2+\Delta\phi^2}$ for a 50 GeV and 100 GeV quark jet
fragmenting in the vacuum (red curves) and a hot medium. Right: Comparison of
the vacuum and medium-induced part of the gluon multiplicity distributions
inside a cone jet of size $R=\Theta_c$. Figures taken from
~\protect\cite{Salgado:2003rv}.}
\end{figure}

A more interesting situation is, however, when the high-$p_t$ particle travels 
through a flowing medium. In this case, the flow introduces a preferred direction
in the medium-induced gluon radiation, producing asymmetric jet shapes
\cite{Armesto:2004pt} (see Fig. 3). These 
asymmetries are, in this way, a measurement of the flow field in a medium.
Interestingly, preliminary results \cite{starprel}
from RHIC on high-$p_t$ two particle
correlations show a strong elongation of the jet-like signal in the
longitudinal direction for central gold-gold collisions. In the spirit of the
effects shown in Fig. 3, this elongation is produced by the strong
longitudinal flow present in these collisions.

\begin{figure}[h]
\begin{center}
\includegraphics[width=5.3cm,angle=-90]{carlosflows.epsi}
\end{center}
\begin{center}
\includegraphics[width=8.5cm]{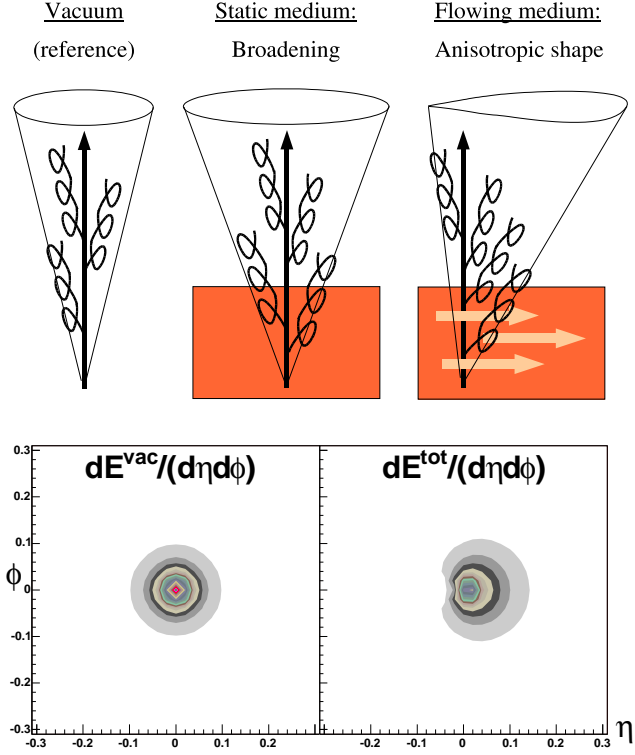}
\end{center}
\caption{Upper part: sketch of the distortion
of the jet energy distribution in the presence of a medium
with or without collective flow. Lower part: calculated
distortion of the jet energy distribution
in the $\eta \times \phi$-plane for a 100 GeV jet. The right
hand-side is for an
average medium-induced radiated energy of 23 GeV and equal
contributions from density and flow effects.
Figures taken from~\protect\cite{Armesto:2004pt}.
}\label{fig3}
\end{figure}

%


\section*{References}
\begin{thebibliography}{99}

\bibitem{exp} See the recent white papers from the RHIC collaborations:
%
  B.B. Back {\it et al.}  [PHOBOS Collaboration], Nucl.\ Phys.\ A{\bf 757}
  (2005) 1;
  I.~Arsene {\it et al.}  [BRAHMS Collaboration],
  Nucl.\ Phys.\ A{\bf 757} (2005) 28;
  K.~Adcox {\it et al.}  [PHENIX Collaboration],
  arXiv:nucl-ex/0410003;
  J.~Adams {\it et al.}  [STAR Collaboration],
  arXiv:nucl-ex/0501009.

\bibitem{reviews} For recent reviews see: 
  R.~Baier, D.~Schiff and B.~G.~Zakharov,
  Ann.\ Rev.\ Nucl.\ Part.\ Sci.\  {\bf 50}, 37 (2000);
%
  M.~Gyulassy, I.~Vitev, X.~N.~Wang and B.~W.~Zhang,
  arXiv:nucl-th/0302077;
%
  A.~Kovner and U.~A.~Wiedemann,
  arXiv:hep-ph/0304151;
%
  C.~A.~Salgado,
  Mod.\ Phys.\ Lett.\ A {\bf 19} (2004) 271;
  P.~Jacobs and X.~N.~Wang,
  Prog.\ Part.\ Nucl.\ Phys.\  {\bf 54} (2005) 443


\bibitem{Baier:1996sk}
  R.~Baier, Y.~L.~Dokshitzer, A.~H.~Mueller, S.~Peigne and D.~Schiff,
  Nucl.\ Phys.\ B {\bf 484} (1997) 265

\bibitem{Baier:2001yt}
  R.~Baier, Y.~L.~Dokshitzer, A.~H.~Mueller and D.~Schiff,
  JHEP {\bf 0109} (2001) 033

\bibitem{Salgado:2003gb}
  C.~A.~Salgado and U.~A.~Wiedemann,
  Phys.\ Rev.\ D {\bf 68} (2003) 014008

\bibitem{Eskola:2004cr}
  K.~J.~Eskola, H.~Honkanen, C.~A.~Salgado and U.~A.~Wiedemann,
  Nucl.\ Phys.\ A {\bf 747} (2005) 511
  
  

\bibitem{Muller:2002fa}
  B.~Muller,
  Phys.\ Rev.\ C {\bf 67} (2003) 061901;
%
  A.~Dainese, C.~Loizides and G.~Paic,
  Eur.\ Phys.\ J.\ C {\bf 38}, 461 (2005)
  
  \bibitem{Armesto:2005iq}
  N.~Armesto, A.~Dainese, C.~A.~Salgado and U.~A.~Wiedemann,
  Phys.\ Rev.\ D {\bf 71} (2005) 054027



\bibitem{hqeloss}
  Y.~L.~Dokshitzer and D.~E.~Kharzeev,
  Phys.\ Lett.\ B {\bf 519}, 199 (2001);
%
  N.~Armesto, C.~A.~Salgado and U.~A.~Wiedemann,
  Phys.\ Rev.\ D {\bf 69} (2004) 114003;
%
  M.~Djordjevic and M.~Gyulassy,
  Nucl.\ Phys.\ A {\bf 733} (2004) 265;
%
  B.~W.~Zhang, E.~Wang and X.~N.~Wang,
  Phys.\ Rev.\ Lett.\  {\bf 93}, 072301 (2004).

\bibitem{Salgado:2003rv}
  C.~A.~Salgado and U.~A.~Wiedemann,
  Phys.\ Rev.\ Lett.\  {\bf 93} (2004) 042301

\bibitem{Armesto:2004pt}
  N.~Armesto, C.~A.~Salgado and U.~A.~Wiedemann,
  Phys.\ Rev.\ Lett.\  {\bf 93} (2004) 242301;
%
  arXiv:hep-ph/0411341.

\bibitem{starprel} F. Wang [STAR Collaboration] nucl-ex/0404010; D. Magestro 
[STAR Collaboration] Proc. of the  International Conference on
Hard and Electromagnetic Probes
of High Energy Nuclear Collisions, Ericeira (Portugal) November 2004.

\end{thebibliography}
\end{document}